\documentstyle[epsf]{mn}

\newif\ifAMStwofonts

\newcommand{\hi}{H\,{\sc i}}

\newcommand{\hubble}{\mbox{$\rm km\, s^{-1}\, Mpc^{-1}$}}
\newcommand{\kms}{\mbox{$\rm km\, s^{-1}$}}
\newcommand{\mhi}{\mbox{$M_{\rm HI}$}}
\newcommand{\mhis}{\mbox{$M^*_{\rm HI}$}}
\newcommand{\icmsq}{\mbox{$\rm cm^{-2}$}}
\newcommand{\ms}{\mbox{$M^*$}}
\newcommand{\magsq}{\mbox{$\rm mag\, arcsec^{-2}$}}
\newcommand{\phis}{\mbox{$\phi^*$}}
\newcommand{\ahiss}{AH{\sc i}SS}
\newcommand{\svmax}{\mbox{$\Sigma(1/V_{\rm max})$}}  
\newcommand{\vmax}{\mbox{$V_{\rm max}$}} 
\newcommand{\etal}{et al.}    
\newcommand{\rhol}{\mbox{$j_{B}$}}  
\newcommand{\lsolb}{\mbox{${\rm L}^B_\odot$}}
\newcommand{\mubi}{\mbox{$\mu_B^{\rm b,i}$}}   
\newcommand{\rhi}{\mbox{$\rho_{\rm HI}$}}   

\newcommand{\ihmpcc}{\mbox{$h_{100}^{3}\, \rm Mpc^{-3}$}}
\newcommand{\ml}{\mbox{$M/L$}}
\newcommand{\rhom}{\mbox{$\rho_{M}$}}  
\newcommand{\lbi}{\mbox{$L_B^{\rm b,i}$}}  
\newcommand{\mbi}{\mbox{$M_B^{\rm b,i}$}}   

\newcommand\mnras{{MNRAS}}%
\newcommand\aap{{A\&A}}%
\newcommand\apj{{ApJ}}%
\newcommand\apjs{{ApJS}}%
\newcommand\araa{{ARA\&A}}%
\newcommand\aj{{AJ}}%
\newcommand\nat{{Nature}}%

\ifoldfss
  \ifCUPmtlplainloaded \else
    \NewTextAlphabet{textbfit} {cmbxti10} {}
    \NewTextAlphabet{textbfss} {cmssbx10} {}
    \NewMathAlphabet{mathbfit} {cmbxti10} {} 
    \NewMathAlphabet{mathbfss} {cmssbx10} {} 
  \fi
  \ifAMStwofonts
    \ifCUPmtlplainloaded \else
      \NewSymbolFont{upmath} {eurm10}
      \NewSymbolFont{AMSa} {msam10}
      \NewMathSymbol{\upi}     {0}{upmath}{19}
      \NewMathSymbol{\umu}     {0}{upmath}{16}
      \NewMathSymbol{\upartial}{0}{upmath}{40}
      \NewMathSymbol{\leqslant}{3}{AMSa}{36}
      \NewMathSymbol{\geqslant}{3}{AMSa}{3E}

      \let\leq=\leqslant 
      \let\geq=\geqslant 
    \fi
  \fi
\fi 

\ifnfssone
  \newmathalphabet{\mathit}
  \addtoversion{normal}{\mathit}{cmr}{m}{it}
  \addtoversion{bold}{\mathit}{cmr}{bx}{it}
  \newmathalphabet{\mathbfit} 
  \addtoversion{normal}{\mathbfit}{cmr}{bx}{it}
  \addtoversion{bold}{\mathbfit}{cmr}{bx}{it}
  \newmathalphabet{\mathbfss} 
  \addtoversion{normal}{\mathbfss}{cmss}{bx}{n}
  \addtoversion{bold}{\mathbfss}{cmss}{bx}{n}
  \ifAMStwofonts
    \ifCUPmtlplainloaded \else
      \UseAMStwoboldmath
      \makeatletter
      \new@mathgroup\upmath@group
      \define@mathgroup\mv@normal\upmath@group{eur}{m}{n}
      \define@mathgroup\mv@bold\upmath@group{eur}{b}{n}
      \edef\UPM{\hexnumber\upmath@group}
      \new@mathgroup\amsa@group
      \define@mathgroup\mv@normal\amsa@group{msa}{m}{n}
      \define@mathgroup\mv@bold\amsa@group{msa}{m}{n}
      \edef\AMSa{\hexnumber\amsa@group}
      \makeatother
      \mathchardef\upi="0\UPM19
      \mathchardef\umu="0\UPM16
      \mathchardef\upartial="0\UPM40
      \mathchardef\leqslant="3\AMSa36
      \mathchardef\geqslant="3\AMSa3E

      \let\leq=\leqslant 
      \let\geq=\geqslant 
    \fi
  \fi
\fi 

\ifnfsstwo
  \DeclareMathAlphabet{\mathbfit}{OT1}{cmr}{bx}{it}
  \SetMathAlphabet\mathbfit{bold}{OT1}{cmr}{bx}{it}
  \DeclareMathAlphabet{\mathbfss}{OT1}{cmss}{bx}{n}
  \SetMathAlphabet\mathbfss{bold}{OT1}{cmss}{bx}{n}
  \ifAMStwofonts
    \ifCUPmtlplainloaded \else
      \DeclareSymbolFont{UPM}{U}{eur}{m}{n}
      \SetSymbolFont{UPM}{bold}{U}{eur}{b}{n}
      \DeclareSymbolFont{AMSa}{U}{msa}{m}{n}
      \DeclareMathSymbol{\upi}{0}{UPM}{"19}
      \DeclareMathSymbol{\umu}{0}{UPM}{"16}
      \DeclareMathSymbol{\upartial}{0}{UPM}{"40}
      \DeclareMathSymbol{\leqslant}{3}{AMSa}{"36}
      \DeclareMathSymbol{\geqslant}{3}{AMSa}{"3E}

      \let\leq=\leqslant 
      \let\geq=\geqslant 
    \fi
  \fi
\fi 

\ifCUPmtlplainloaded \else
  \ifAMStwofonts \else 
    \def\upi{\pi}
    \def\umu{\mu}
    \def\upartial{\partial}
  \fi
\fi

 \title[The Luminosity Function and Surface Brightness Distribution of
H\,{\normalsize\it I} Selected Galaxies]{The Luminosity Function and
Surface Brightness Distribution of H\,{\Large\bf I} Selected Galaxies}
 \author[M. A. Zwaan, F. H. Briggs and D. Sprayberry]
       {Martin A. Zwaan,$^{1,2}$\thanks{email: mazwaan@unimelb.edu.au} 
	Frank H. Briggs$^2$ and 
	David Sprayberry$^3$ \\
 	$^1$  	Astrophysics Group,
        	School of Physics,
        	University of Melbourne,
        	Victoria 3010,
        	Australia\\
        $^2$ 	Kapteyn Astronomical Institute, 
		P.O. Box 800, 
		9700 AV Groningen, 
		The Netherlands\\
        $^3$ 	W.~M. Keck Observatory, 
		65-1120 Mamalahoa Highway,
		Kamuela,
		Hawaii 96743\\
	}

\date{Accepted ...
      Received ...}

\pagerange{\pageref{firstpage}--\pageref{lastpage}}
\pubyear{0000}

\begin{document}

\maketitle

\label{firstpage}

 \begin{abstract}
 We measure the $z=0$ $B$-band optical luminosity function (LF) for galaxies
selected in a blind \hi\ survey.   
The total LF of the \hi\ selected sample is
flat, with Schechter parameters $\ms=-19.38_{-0.62}^{+1.02} + 5\log
h_{100}$ mag and $\alpha=-1.03_{-0.15}^{+0.25}$, in good agreement with
LFs of optically selected late-type galaxies.  Bivariate distribution functions of
several galaxy parameters show that the \hi\ density in the local
Universe is more widely spread over galaxies of different size, central surface
brightness, and luminosity than is the optical luminosity density.
The number density of very
low surface brightness ($>24.0$~\magsq) gas-rich galaxies is
considerably lower than that found in optical surveys designed to detect
dim galaxies.  This suggests that only a part of the population of LSB
galaxies is gas rich and that the rest must be gas poor.  However, we
show that this gas-poor population must be cosmologically insignificant
in baryon content.  The contribution of gas-rich LSB galaxies
($>23.0$~\magsq) to the local cosmological gas and luminosity density is
modest ($18_{-5}^{+6}$ and $5_{-2}^{+2}$ per cent respectively); their
contribution to $\Omega_{\rm matter}$ is not well-determined, but
probably $<$ 11 per cent.  These values are in excellent agreement with the
low redshift results from the Hubble Deep Field. 
 \end{abstract}

\begin{keywords}
galaxies: luminosity function, mass function --
galaxies: statistics --
galaxies: fundamental parameters
\end{keywords}

\section{Introduction}
 Understanding galaxy evolution requires well-determined local
benchmarks.  One of the most fundamental of these is the field galaxy
luminosity function, the shape of which should be predicted by any
reliable galaxy formation theory.  In principle, the shape of the
luminosity function is related to the power spectrum of primordial
density fluctuations and complex processes such as gas cooling, star
formation and feedback to the interstellar medium as well as the
behaviour of dark matter as it undergoes gravitational collapse and
merging in galaxy halos (see e.g., Cole et al.  2000 for a recent
review).  Reference points in the local Universe will help in developing
a full understanding of these processes.  Another motive for
determination of the local luminosity function, is the problem of the
faint blue galaxies.  The normalisation of the $z=0$ luminosity function
seems to be too low to be reconciled with no-evolution predictions based
on intermediate redshift ($z\sim 1$) surveys (Ellis 1997; Broadhurst,
Ellis \& Shanks 1988; Koo \& Kron 1992), but to quantify this problem a
reliable measurement of the faint end slope is essential. 

 The last few years have seen a proliferation of published luminosity
functions from optical redshift surveys of the local ($z<0.2$) Universe
(see e.g., Zucca et al.  1997; Ratcliffe et al.  1998; Folkes et al. 
1999; and Blanton et al.  (2001) for some recent examples).  These
surveys systematically produce samples of $10^4$ galaxies and are able
to determine the luminosity function down to absolute magnitude limits
of $M_B=-14$.  However, considerable uncertainty remains about the exact
shape and normalisation of the luminosity function.  Especially the
faint-end slope for the dwarf galaxies ($M_B>-18$) is practically
unconstrained (see discussion in Driver \& Phillipps 1996). 

 A potential cause of the uncertainty in low $z$ galaxy counts is the
surface brightness selection effect (Disney 1976; Disney \& Phillips
1987).  Sprayberry et al.  (1997) specifically searched for the low
surface brightness (LSB) galaxies in the APM survey (Impey et al.
1996), and concluded that including LSB galaxies in the low $z$ census 
steepens the field luminosity function, but still does not close the gap
between number counts at moderate redshift and $z=0$.

 The discussion on LSB galaxies ties in directly with another important
benchmark at $z=0$: the distribution function of optical surface
brightnesses.  Based on photometry of 36 nearby spiral and S0 galaxies,
Freeman (1970) concluded that $\sim 80$ per cent have a $B$ band central
surface brightness $\mu_B(0)$ in the range $21.65 \pm 0.30$~\magsq.  The
eight deviant galaxies consisted of one dwarf irregular LSB galaxy
($\mu_B(0)=23.7~\magsq$) and seven brighter galaxies of various
morphological type.  The majority opinion at the present moment seems to be
that the distribution function is flat (McGaugh 1996; de Jong 1996;
Dalcanton et al.  1997b; O'Neil \& Bothun 2000; de Jong \& Lacey 2000;
Blanton et al.  2001; Cross et al.  2001), although Sprayberry et al.
(1996) found a distribution function that peaks at $\sim 22$~\magsq.
Tully \& Verheijen
(1997) have a dissenting view and present evidence for bimodality in the
distribution of near-infrared surface brightnesses in the Ursa Major
Cluster.  This view has been contested by Bell \& de Blok (2000) who
claim that the data set is insufficient to establish the presence of a
bimodal surface brightness distribution. 

New insight in both issues can be obtained by selecting galaxies via a
method that is free from optical selection effects.  In this paper we
measure for the first time the optical luminosity function and surface
brightness distribution function of \hi\ selected galaxies.  This sample
is the result of the Arecibo \hi\ Strip Survey, a blind strip survey in
the 21cm line.  We stress that this sample is small (60 members)
compared to those produced by modern redshift surveys, and large
statistical errors are therefore unavoidable.  This work should be
regarded as the first step toward measuring these functions for \hi\
selected galaxy samples.  Much larger galaxy samples will be available
in the near future (e.g.  HIPASS, Staveley-Smith et al.  1996), and the
measurements of optical luminosity functions, surface brightness
functions and bivariate distributions will greatly improve. 

We organise this paper as follows.  First, in
section~\ref{datashort.sec}, we briefly describe the sample.  In
section~\ref{lfs.sec} we present the optical luminosity function of this
\hi\ selected galaxy sample, and discuss the distribution of luminosity
density and \hi\ gas density among different galaxies.  In
section~\ref{sbds.sec} the surface brightness distribution function and
the contribution of LSB galaxies to the mass density of the local
Universe are discussed.  Bivariate distribution functions of various
galaxy parameters are presented in section~\ref{bivar.sec}.  Finally, in
section~\ref{conc_opt.sec} we summarise the conclusions.  Throughout
this paper we use $H_0=100\,h_{100} \hubble$ for calculating distance
dependent quantities. 

\section{The data} \label{datashort.sec}
 The sample of galaxies used here to measure distribution functions of
optical luminosity and central surface brightness is selected in the
21cm line, and it is therefore free from selection effects related to
optical surface brightness.  The sample is the result of the Arecibo
\hi\ Strip Survey (\ahiss), a blind extragalactic \hi\ survey consisting
of two strips of constant declination, together covering approximately
65 square degrees of sky over a depth of $cz=7500$~\kms.  The limiting
column density for the central 3 arcmin wide strip (corresponding to the
main beam of the Arecibo Telescope) was $\approx 10^{18} \rm cm^{-2}$
($5\sigma$) per resolution element of 16~\kms for gas filling the
telescope beam.  This sensitivity is unmatched by any other blind \hi\
survey to date.  Low resolution 21cm aperture synthesis observations of
the \ahiss\ sample of 66 galaxies have been obtained with the NRAO Very
Large Array (VLA).  Details of the Arecibo survey and the VLA
observations are described by Sorar (1994) and Zwaan et al.  (1997). 

Optical observations were confined to sources at Galactic latitudes
$|b|>10^\circ$ to avoid severe Galactic extinction and confusion of
foreground stars.  This reduces the total number of accessible sources
to 61.  The optical data were obtained at the Isaac Newton Telescope of
the Observatorio del Roqu\'{e} de los Muchachos on the island of La
Palma, Spain.  The data collection was spread over four observing runs
during the period 1995 October through 1997 March.  Images were recorded
at the Prime Focus camera with a thinned Tektronix $1024^2$ pixel CCD. 
The Tektronix CCD has 24 $\mu$m pixels, which give an image scale of
0.59$''$ per pixel at prime focus.  All images were taken through a
standard Harris $B$ filter.  Flatfields were taken in the twilight, and
the residual background variations after flatfielding are typically $<
1$ per cent of the sky level. 
 The photometric calibration was done by observing standard stars at
several airmasses each night, and is accurate to 0.13 mag.

Total galaxy magnitudes were determined using aperture photometry on  
the reduced images.  Correct aperture sizes were found using a
curve-of-growth algorithm: aperture photometry was performed at a series
of aperture radii, increasing in 1 arcsecond steps, until the integrated
magnitudes levelled out at an asymptotic maximum.  The first radius at
which this maximum (brightest) integrated magnitude was reached was then
chosen as the correct aperture size.  
 Central surface brightnesses and disk scale lengths were determined by
fitting exponential disk models to the azimuthically averaged radial
surface brightness profiles.  The centres of the galaxies were usually
taken to be the maximum of the light distribution.  Strong central
concentrations were excluded from the exponential fits. The data for 
galaxy A44 turned out to be not usable, which leaves us with 
a total number of 60 galaxies for our analysis.

We have chosen to apply the internal extinction correction proposed by
Tully \etal\ (1998) which is a function of absolute magnitude.  The
extinction correction can be written as $A^{\rm i}=\gamma \log(a/b)$,
where $\gamma=-0.35(15.1+\mbi)$.  Since for our data set the central
surface brightness is well correlated with absolute luminosity, the
extinction correction implies a low correction for LSB galaxies and a
higher correction for HSB galaxies. Galactic extinction corrections
have been applied using the reddening maps of Burstein \& Heiles (1982) 
and assuming that $A(B) = 4.1\, E(B - V)$.

\section{Luminosity functions} \label{lfs.sec}
 The luminosity function (LF) of galaxies is defined as the number of
galaxies per cubic Mpc in a luminosity interval $dM$ centred at
magnitude $M$. The interval $dM$ is generally taken to be 1 mag. The  
most used parameterisation of the luminosity function is the Schechter
(1976) function defined by
 \begin{eqnarray}
 \phi(M)dM = 0.4\,{\rm ln}\, 10\,\phis\,[10^{0.4(M^*-M)}]^{1+\alpha} 
 \times\nonumber\\ 
\exp[-10^{0.4(M^*-M)}]\,dM,
 \end{eqnarray}
where $\alpha$ is the faint-end slope, \phis\ is the normalisation  
factor and \ms\ is the characteristic absolute magnitude
that defines the boundary between the exponential and power-law part.

\subsection{Methods}
 Many different galaxy luminosity function estimators can be found in
the literature.  In Zwaan et al.  (1997) we discuss different luminosity
function estimators and conclude that the \svmax\ method is the
preferred way to determine mass functions and luminosity functions for
our sample.  For this sample, we demonstrated that the determination of
the \hi\ mass function with the \svmax\ method is not very sensitive to density
fluctuations due to large scale structure.  The \svmax\ method consists
of summing the reciprocals of the volumes corresponding to the maximum
distances at which galaxies could be seen and still remain within the
sample.  Summing these values per bin in \hi\ mass or absolute magnitude
immediately gives the binned \hi\ mass function or optical luminosity
function.  The advantages of the \svmax\ method are that it is
automatically normalised and non-parametric; it recovers the amplitude
and the shape of the luminosity function simultaneously, without using
the Schechter function as an assumption about the intrinsic shape.  An
overview of the different galaxy luminosity function estimators is given
by Willmer (1997), who tests the validity of different methods by means
of Monte-Carlo simulations.  Careful examination of his tables shows
that the \svmax\ method (with binning in magnitudes) recovers the input
luminosity function satisfactorily, and equally well as the more
conventional parameterised maximum likelihood method (Sandage, Tammann
\& Yahil 1979) or the Stepwise Maximum Likelihood Method (SWLM,
Efstathiou, Ellis \& Peterson 1988).  Supported by this, we choose to
apply the \svmax\ method to evaluate the optical luminosity function. 
The details of the determination of the values of \vmax\ are described
in Zwaan et al.  (1997) and will not be repeated here. Note that the
values of \vmax\ are derived from the original Arecibo \hi\
survey parameters and are independent from the optical data.

\subsection{Results}
\begin{figure}
\epsfxsize=8.8cm \epsfbox[60 195 560 650]{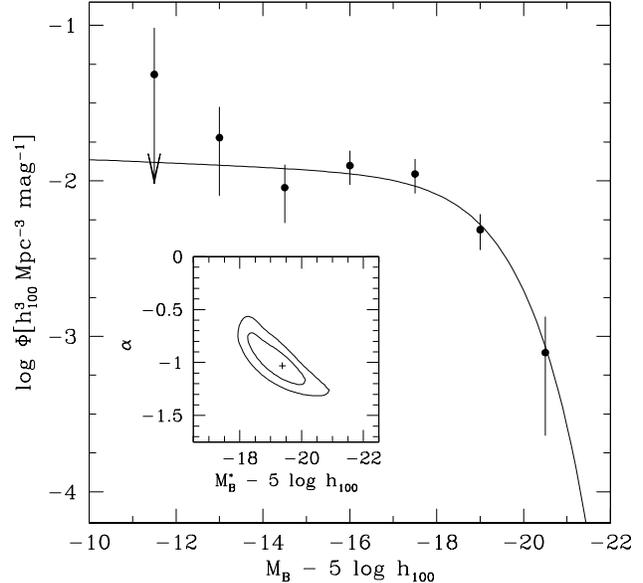}
 \caption{Luminosity function for \hi\ selected galaxies.  The points
were determined using the \vmax\ method, the errorbars are 1$\sigma$
uncertainties from Poisson statistics.  The line is the best fit
Schechter function with parameters:  $\alpha=-1.03_{-0.15}^{+0.25}$,
$\ms=-19.38_{-0.62}^{+1.02} + 5\log h_{100}$ mag
and $\phis=(1.15 \pm 0.40)~\times
10^{-2}\,h_{100}^{-3} \rm Mpc^{-3}$.  The inset shows the 1$\sigma$ and
2$\sigma$ joint two-parameter confidence levels for $\alpha$ and \ms.}
\label{lf.fig}
\end{figure}

 The resulting luminosity function $\phi(M_B)$ is shown in
Fig.~\ref{lf.fig} as solid dots with 1$\sigma$ errorbars.  The data
are binned per 1.5 mag in order to obtain a reasonable number of
galaxies per bin, but scaled in such a way that $\phi$ represents number
densities per magnitude bins.  Furthermore, the data are multiplied by a
factor 66/60 to account for the galaxies for which no optical
information is available.  To enable direct comparison with published
luminosity functions, we choose to use the absolute magnitudes
uncorrected for opacity effects in the galactic disk (see Leroy \&
Portilla 1998 for a discussion on the influence of optical depth   
effects on the shape of the luminosity function).  The line indicates
the best fit Schechter function which is determined by minimising
$\chi^2$ for the expected number of galaxies per bin.  The uncertainties
in the best fit are indicated in the inset that shows the 1$\sigma$ and
2$\sigma$ error contours of the $\chi^2$ fit for $\alpha$ and \ms\
fitted jointly.  As is usually the case in these fits, the parameters  
$\alpha$ and \ms\ are strongly correlated in the sense that steeper  
faint end slopes imply brighter values of \ms.  The best fit Schechter
parameters are found to be
 $\alpha=-1.03_{-0.15}^{+0.25}$,
 $\ms=-19.38_{-0.62}^{+1.02} + 5\log h_{100}$ mag and
 $\phis=(1.15 \pm 0.40)~\times 10^{-2}\,\,h_{100}^{-3}\rm
Mpc^{-3}$, where the quoted errors are 1$\sigma$ one-parameter  
uncertainties.  
The uncertainties given here are solely the result of   
counting statistics, and therefore may understate the true
uncertainties.  Measurement errors in the parameters that define \vmax\
and measurement errors in $M_B$ also contribute to the uncertainties,
but these are relatively small compared to the Poisson errors for this
small sample.

The parameterisation in the form of a Schechter function is a
satisfactory representation of luminosity function of the \ahiss\
galaxies.  However, due to the small number of galaxies in the low
luminosity bins, the value of the faint end slope $\alpha$ is poorly   
constrained.  Especially for magnitudes fainter than $M_B=-14$, the
slope of the LF is almost unconstrained.  There is no need for a
modification of the LF, such as the Schechter function plus a power law,
proposed by Sprayberry \etal\ (1997) for his sample of LSB galaxies,  
although our present sample does not rule out this extra component.
We note that our measured LF parameters are in excellent agreement with
those from a preliminary analysis of the HIPASS survey (Marquarding 2000).

\subsection{Comparison with optical determinations of the LF}
 It is interesting to compare the luminosity function for \hi\ selected
galaxies to luminosity functions of optically selected galaxies.
Recently, there has been much interest in steep faint-end slopes of the 
luminosity function, and the galaxies responsible for this steep part 1)
are found to be of late morphological type (e.g., Marzke et al.  1998),
2) show strong emission lines indicative of active star formation (e.g.,
Zucca et al.  1997), and 3) have blue colours (Lin et al.  1999).  These
are the types of galaxies that are expected to contain high fractions of
\hi, and therefore should be represented in the \ahiss\ sample.

A vast number of luminosity functions based on optical redshift redshift
surveys, is available in the literature.  All these surveys contain
typically a few thousand galaxies.  When making a comparison with our  
luminosity function for \hi\ selected galaxies, we will concentrate on
those studies which have made a specific distinction between late and
early type galaxies, or star forming and quiescent galaxies.  We
consider: the Stromlo-APM redshift survey (APM, Loveday et al.  1992),
the Center for Astrophysics redshift survey (CfA, Marzke et al.  1994),
the ESO Slice Project (ESP, Zucca at al.  1997), the Las Campanas
Redshift Survey (LCRS, Lin et al.  1996), the Autofib Redshift Survey
(ARS, Heyl et al.  1997), the Second Southern Sky Redshift Survey
(SSRS2, Marzke et al.  1998), the CNOC Field Galaxy Redshift Survey
(CNOC2, Lin et al.  1999) and preliminary results from the 2dF survey   
(Folkes et al. 1999).

\begin{figure}
\epsfxsize=8.8cm \epsfbox[60 195 560 650]{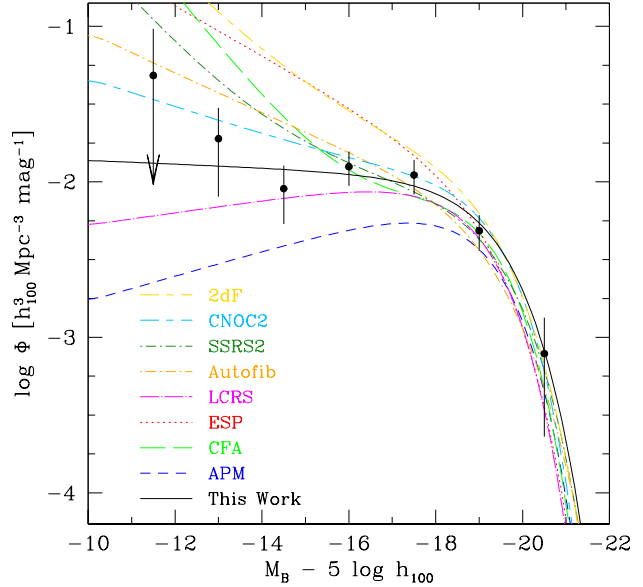}
 \caption{Luminosity functions for late type galaxies.
The points are the same as in Fig.~\ref{lf.fig}. The lines show the
luminosity functions from several recent redshift surveys. The details
are given in Table~\ref{tableLF.tab}. Some of these are the summations
of several luminosity functions for different types.}
\label{lfslate.fig}
\end{figure}

Table~\ref{tableLF.tab} summarises the Schechter parameters of `late
type' luminosity functions of these surveys.  In
Fig.~\ref{lfslate.fig} these functions are represented, together with
our measured points and the best fit Schechter function.  The functions
have been transformed to the $B$ filter using the conversions $M_B=   
M_Z-0.45$ for the CfA, $b-r=1.1$ for the LCRS, and $M_B-M_{b J}=0.24$,
and all functions are recalculated for $H_0=100\,h_{100}\, \hubble$.

 \begin{table*}
 \begin{minipage}{12.2cm}
 \caption[]{Comparison of Luminosity Functions for Late Type Galaxies}
 \label{tableLF.tab}
 \begin{tabular}{llcccc}
 \hline
 \noalign{\smallskip}
{Sample} &
{Selection} &
{$\alpha$} &
{$M^* - 5 \log h_{100}$} &
{$\phi^*$$^{\rm a}$} &
{$\rho_{L}$$^{\rm b}$}   \\
\noalign{\smallskip}
AH{\sc i}SS 	  
		  & H~{\sc i} selected 	     &$ -1.03      $&$ -19.38	$& 11.5		 & $10.3 \pm 2.0$	\\
APM (Loveday et al. 1992)& Sp/Irr		     &$ -0.80      $&$ -19.16	$& 10 		 &  6.7 	\\
CfA (Marzke et al. 1994) & Sa-Sb   		     &$ -0.58      $&$ -18.93^{\rm c}\!\!\! 	$& 8.7  	 &  8.9	     	\\
		  & Sc-Sd   		     &$ -0.96      $&$ -19.02^{\rm c}\!\!\! 	$& 4.4  	 & 	   	\\
		  & Sm-Im   		     &$ -1.87      $&$ -19.00^{\rm c}\!\!\! 	$& 0.6  	 & 	   	\\
ESP (Zucca et al. 1997)  & Emission lines 	     &$ -1.40      $&$ -19.23 	$& 10	 	 & 11.4     	\\
LCRS (Lin et al. 1997)   & $3727\,W_\lambda \geq 5$ \AA 	
                                             &$ -0.90      $&$ -18.93^{\rm d}\!\!\!$& 13	 &  7.7 	\\
Autofib (Heyl et al. 1997)
  	          & Sab 		     &$ -0.99      $&$ -19.76 	$& 2.19 	 &  8.4		\\
                  & Sbc                      &$ -1.25      $&$ -19.16   $& 2.80         & 	   	\\
                  & Scd                      &$ -1.37      $&$ -18.96   $& 3.01         & 	   	\\
		  & Sdm     		     &$ -1.36      $&$ -18.76 	$& 0.50 	 & 	   	\\
SSRS2 (Marzke et al. 1998)
		  & Spirals		     &$ -1.11	   $&$ -19.43   $& 8.0 		& 9.2		\\
		  & Irr/Pec		     &$ -1.81	   $&$ -19.78   $& 0.2		&		\\
CNOC2 (Lin et al. 1999)$^{\rm e}$  		  
	          & Intermediate type	     &$ -0.53	   $&$ -18.97	$& 9.0		 & 11.1		\\
                  & Late type		     &$ -1.23	   $&$ -19.07 	$& 7.2		 &		\\
2dF (Folkes et al. 1999)  & Sab			     &$ -0.86	   $&$ -19.44   $& 3.9		& 14.0 \\
	       	  & Sbc			     &$ -0.99 	   $&$ -19.14   $& 5.3&\\
		  & Scd			     &$ -1.21	   $&$ -18.76   $& 6.5&\\
		  & Sdm-Im		     &$ -1.73	   $&$ -18.78   $& 2.1&\\
 \noalign{\smallskip} 
 \hline
 \end{tabular}
 \medskip
 \begin{list}{}{}
\item $^{\mathrm{a}}$~Units are $10^{-3}\,h_{100}^3\rm\,Mpc^{-3}$
\item $^{\mathrm{b}}$~Units are $10^7\,h_{100}\,\lsolb\rm\,Mpc^{-3}$
\item $^{\mathrm{c}}$~$B-M_Z=-0.21$
\item $^{\mathrm{d}}$~$B-R=1.1$
\item $^{\mathrm{e}}$~Values extrapolated to $z=0$
 \end{list}
\end{minipage}
\end{table*}

The direct comparison of these luminosity functions is rather naive for
a number of reasons.  Firstly, different optical wave bands have been
used in the selection of these galaxies.  This effect can be corrected
for by applying a magnitude correction, but this is most certainly a 
oversimplification of the problem.  The use of different wave bands does
not only have an influence on the luminosity of the selected galaxies  
but surely also on the morphological classifications.
Secondly, the separation between
late and early type galaxies has been made in different ways for each 
sample.  In the CfA sample a detailed separation between morphological
types has been made on the basis of the galaxies' appearances on Palomar
Sky Survey.  Also the SSRS2 and the APM samples have been classified by
visual inspection.  The CNOC2 data is split into different populations
using colour information of the galaxies.  For the 2dF and Autofib
surveys spectral information has been used to make the classifications.
The selection criteria for the ESP and the LCRS samples has been the
occurrence of emission lines in the spectra.  In the LCRS sample a     
distinction has been made on basis of the criterion of [O{\sc ii}]
$3727\,W_\lambda \geq 5$ \AA, in the EPS sample the selection was simply
based on the detection the [O{\sc ii}] line.

With these restrictions in mind, we can compare the different luminosity
functions for optically selected galaxies with \hi\ selected galaxies. 
What is particularly striking is that the values of the faint-end slope
span a wide range from $-0.80$ for the APM survey to $\sim -1.50$ for
the ESP and 2dF surveys.  Even for surveys that use comparable methods 
for classifying their different galaxy population, the differences in
faint-end slope can be large.  Evidently, the shape of the luminosity
distribution of late type galaxies is still ill-constrained.  On the  
other hand, the normalisation and the value that defines the knee are 
quite similar for all surveys; all functions cross approximately the
same point at $M_B \approx -19 + 5 \log h_{100}$.  The luminosity
function for the \ahiss\ falls in between those of the optical samples.
We therefore conclude that our estimate of the luminosity function is in
good agreement with that of optically selected samples.
Furthermore, there is no new population selected by \hi\ surveys that
adds significantly to the galaxy populations identified through
optical surveys.

\subsection{Luminosity density of gas-rich galaxies}
 A more fundamental parameter is the luminosity density, the integrated 
light from the whole population of galaxies.  As is discussed by Lilly
et al.  (1996), this parameter is in principle less dependent on the
details of galaxy evolution than the luminosity function.
 The integral luminosity density of late type galaxies can be determined
by integrating the Schechter luminosity function weighted by luminosity, which   
gives $\rhol=\phis\,L_{B}^*\,\Gamma(2+\alpha)$, where $\Gamma$ is the
Euler gamma function.  The values of \rhol\ for late-type galaxies as  
determined by the different optical surveys is given in the last column
of Table~\ref{tableLF.tab}.  It is remarkable that all values of \rhol\
are within $\sim 1.5\sigma$ from the value determined from the \ahiss.
A notable exception is the 2dF survey that finds a value 60 per cent higher  
than the mean of the other surveys.  Folkes et al.  (1999) note that the
2dF results are preliminary, and that corrections for completeness,
clustering, and Malmquist-bias have not been applied yet. It remains to
be seen whether the final 2dF results will remain in excess of the
\ahiss\
estimate of \rhol.
 A preliminary result from the SDSS (Blanton et al.  2001) has a
substantially higher optical luminosity density than other recent
surveys, but this increase appears to arise in a their photometric
evaluation of each galaxy's luminosity rather than an increase in the
number density of objects. 

The mean value of $\rhol$ of all optically selected late-type 
galaxy samples is
$9.7 \times 10^7\,h_{100}\,\lsolb\rm\,Mpc^{-3}$, while that for the
\ahiss\ sample is $(10.3\pm 2.0) \times
10^7\,h_{100}\,\lsolb\rm\,Mpc^{-3}$.  This latter value translates to
$\rhol=(3.4\pm 0.7) \times
10^{19}\,h^{-2}_{100}\rm\,W\,Hz^{-1}\,Mpc^{-3}$, using the conversion of
Lilly et al.  (1996).  This is approximately 50 per cent of the integral
luminosity density of the local Universe as measured by the most recent
optical redshift surveys using isophotal magnitudes 
(Folkes et al.  1999, Blanton et al.  2001). Blanton et al. (2001) find
that \rhol\ increases significantly if extrapolated magnitudes are used.

\subsection{Luminosity and \hi\ mass distributions for     
different morphological types}
 A more detailed view of the relative importance of different
morphological types to the \hi\ and luminosity density can be made be 
transforming luminosity functions into \hi\ mass functions, assuming  
correlations between \hi\ mass and optical luminosity.  Rao \& Briggs   
(1993) used this method to determine the \hi\ mass function and
$\Omega_{\rm HI}$ based on at that time available luminosity functions.
They showed that by adopting the
relation $\log \mhi = a - bM_B$ between \hi\ mass and optical
luminosity, the \hi\ mass function can be written as
\begin{eqnarray}
\Theta (\mhi) d(\mhi)=\frac{0.4}{b}
 \,\phis\, (\frac{\mhi}{\mhis})^{(\alpha+1)\frac{0.4}{b}-1} \times\nonumber \\
 \exp -(\frac{\mhi}{\mhis})^{\frac{0.4}{b}}\,d(\frac{\mhi}{\mhis}),
\end{eqnarray}
where $\log \mhis = a - bM_B^*$, and $\alpha$ and \phis\ are the 
Schechter parameters of the luminosity functions.

Here we update the calculations by Rao \& Briggs (1993) with more recent
luminosity function parameters, and test if the results are in agreement
with our measurements.  For completeness, we present all possible ways 
of plotting the number density, the \hi\ density, and the luminosity   
density as a function of absolute magnitude and \hi\ mass.  We adopt the
Marzke et al.  (1998) luminosity functions for different morphological
types, and we fit linear regression lines to \mhi\ vs.  $M_B$ taken from
the Nearby Galaxy Catalog (Tully 1988) to find the values of $a$ and
$b$.

\begin{figure}
\epsfxsize=8.8cm \epsfbox[40 170 430 700]{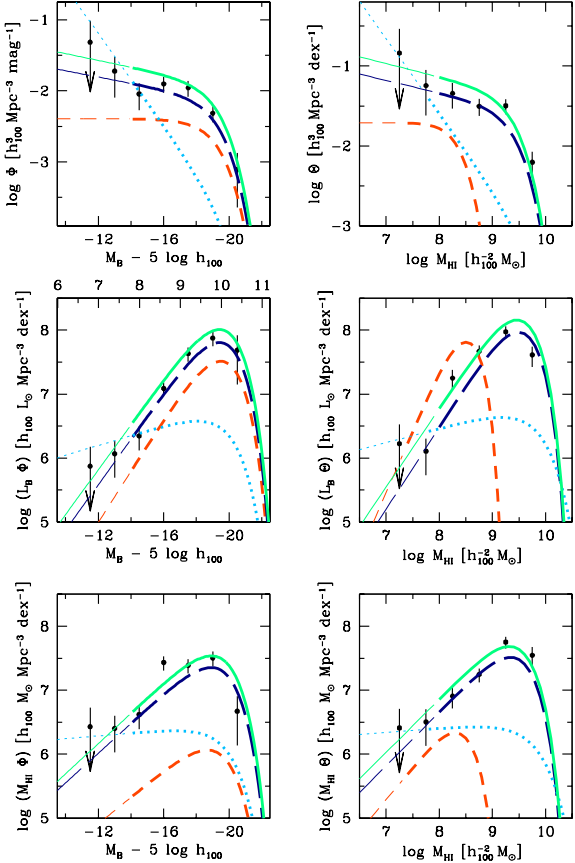}
 \caption{Luminosity and \hi\ density functions for different
morphological types.  The points are from the \ahiss\ and are the result
of the 1/\vmax\ method.  The lines are converted functions for different
morphological types, using Marzke et al.  (1998) luminosity functions,
and \hi\ to $M_B$ relations from the Nearby Galaxy Catalog (Tully 1988). 
The solid grey lines are for all galaxy types, the black dashed lines
for spirals, the grey dashed lines for E and S0, and the grey dotted
lines for Irr/Pec.  The thin parts of each line are extrapolations
beyond the confidence levels set by Marzke et al.  (1998).  {\em Upper
left panel\/}: Luminosity function.  {\em Upper right panel\/}: \hi\
mass function.  {\em Middle left panel\/}: Luminosity density as a
function of $M_B$.  {\em Middle right panel\/}: Luminosity density as a
function of \mhi.  {\em Lower left panel\/}: \hi\ density as a function
of $M_B$.  {\em Lower right panel\/}: \hi\ density as a function of
\mhi.} 
\label{lfmosa.fig} 
\end{figure}

 Fig.~\ref{lfmosa.fig} shows the results.  The solid grey lines are
for all galaxy types, the black dashed lines are for spirals, the grey  
dashed lines for E and S0 types, and the grey dotted line for Irr/Pec   
types.  The thin parts of each line are extrapolations beyond the
confidence levels set by Marzke et al.  (1998).  The solid points plus 
errorbars are different representations of the \ahiss\ data.  The
results are basically the same as what Rao \& Briggs (1993) found.  We
show here that the optical luminosity functions, combined with
conversion factors from $M_B$ to \mhi, give excellent fits to our data.
The \hi\ density distribution matches the converted luminosity functions
for all galaxy types summed, and the luminosity distribution is fitted
satisfactorily with spiral and irregular population.  It is no surprise
that the luminosity density from ellipticals exceeds the measured values
from the \ahiss, since these objects are not selected by \hi\ surveys. 
This is especially clear in the middle right panel.
 
The integral \hi\ density can be determined from a optical luminosity 
functions via
\begin{equation}
\rhi = \int_{-\infty}^{+\infty} \Phi(L)\, \mhi\, dL
     = \phis\,10^{a-b\,M^*}\, \Gamma(1+\alpha+2.5\,b).
 \end{equation}
 If we apply this to our adopted luminosity functions we find that
spirals make up 62 per cent of the \hi\ gas density, Irr and Pec types
contribute 35 per cent, and E and S0 types only 3 per cent.  
Natarajan \& Pettini
(1997) apply this same method to measurements of the luminosity function
at higher redshift in order to chart the evolution of the cosmic gas
content between $z=1$ and $z=0$.  The viability of this result is
unclear since the amount of evolution of the $M_{\rm HI}/L$ ratio of
galaxies is presently unknown.  Future deep \hi\ surveys at redshifts 
$z>0$ are required to constrain the $M_{\rm HI}/L$ evolution.

\section{Contribution of LSB galaxies to the cosmic mass budget}
\label{sbds.sec}
 More than two decades after the seminal paper by Disney (1976) who
defined the potential selection effects against LSB galaxies, the debate
on the cosmological significance of LSB galaxies is still open.  The
\ahiss\ sample, which is not biased by the sky background,  makes a
valuable contribution to the discussion of the cosmological significance
of LSB galaxies.
     
 \subsection{The surface brightness distribution function}
\begin{figure}
\epsfxsize=8.8cm \epsfbox[60 195 560 650]{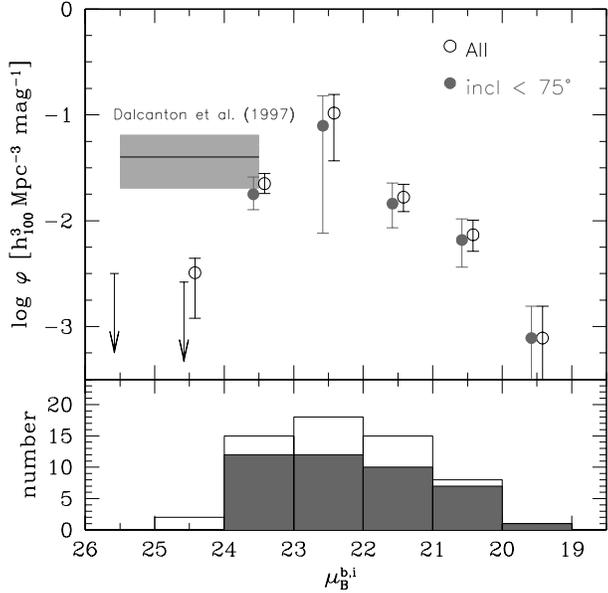}
 \caption{{\em Bottom panel\/}: Distribution of surface brightnesses in
the \ahiss\ sample.  The grey histogram is for those galaxies with $i
\leq 75^\circ$ and is embedded in the histogram for all galaxies. 
 {\em Top panel\/}: Volume corrected distribution of surface
brightnesses.  The open and solid symbols have the same meaning as in
the bottom panel.  For clarity, the points are slightly offset
horizontally.  Errorbars indicate 68 per cent confidence levels.  Arrows
denote 95 per cent confidence upper limits.  The space density of
optically selected LSB galaxies determined by Dalcanton et al.  (1997b)
is indicated by a light grey box, and corresponds to 90 per cent
confidence levels.} 
\label{sbhist.fig}
\end{figure}

 The bottom panel of Fig.~\ref{sbhist.fig} shows a histogram of the
distribution of $B$-band surface brightnesses for the \ahiss\ galaxies.
These surface brightnesses are 
corrected for dust extinction
following the formalism described by Tully et al.  (1998).  The unshaded
histogram shows the distribution for the full set of \ahiss\ galaxies,
and the grey histograms show the distribution for the subset of \ahiss\
galaxies with inclinations $i \leq 75^\circ$ for which the corrections
to face-on values are modest.  The galaxies with high inclinations ($i >
75^\circ$) do not appear from their optical images to be very low
surface brightness: they often exhibit bright central condensations and
strong dust lanes, both features not normally found in extreme LSB
galaxies (see McGaugh, Schombert, \& Bothun 1995).  It therefore seems
possible that the true face-on surface brightness of these disks is
brighter than those given by either the Tully et al. 
(1998) prescription or the
assumption that the disks are fully transparent.

The top panel of Fig.~\ref{sbhist.fig} shows the volume-corrected
surface brightness distribution function of \ahiss\ galaxies.  This
function is determined by summing values of $1/\vmax$ per 1 mag bins of
surface brightness.  The errorbars indicate 68 per cent confidence
levels and are determined from 100 bootstrap re-sample realizations of
the data.  The hollow symbols show the distribution for the complete set
of \ahiss\ galaxies, and the solid symbols are limited to those \ahiss\
galaxies with $i \leq 75^\circ$.  Note that the distribution function
resembles the one found by Sprayberry et al.  (1996) based on the APM
survey. 

\subsection{A cutoff in surface brightness?} \label{cutoff.sec}
 As Fig.~\ref{sbhist.fig} shows, the \ahiss\ detected no galaxies with
reliably determined face-on central surface brightnesses fainter than  
$\mubi = 24$~\magsq.  Even among the highly inclined galaxies with large
(and possibly unreliable) corrections to face-on values, there are no 
galaxies with $\mu_B \ga 25~\magsq$. 
 The statistical significance of this results depends on the assumptions 
we make about the detectability of very LSB systems.  We can make an
estimate by calculating the average value of \vmax\ for the different
surface brightness bins.  We find that \vmax\ is mildly correlated with
\mubi: dimmer galaxies can on average be detected over smaller volumes.
This correlation arises because decreasing surface brightness correlates
with decreasing total \hi\ mass, and the sample selection is based on
\hi\ flux.  If we extrapolate the $\mubi-\vmax$ correlation to the
surface brightness bins in which we have no detections, we find that
$\langle \vmax \rangle$ would be
 $1150\, h_{100}^{-3}\, \rm Mpc^3$ for the $24 - 25~\magsq$ bin, and
 $950\, h_{100}^{-3}\, \rm Mpc^3$ for the $25 - 26~\magsq$ bin.
The probability $p_k$ of finding $k$ objects when
the mean is $n$, is given by the Poisson distribution:
 $
p_k = {e^{-n} n^k}/{k!}.
 $
The mean number of detected objects per mag is given by
$\varphi(\mu)V_{\rm max}(\mu)$, where $\varphi(\mu)$ is the space
density of objects as a function of surface brightness. Hence,
the probability of finding zero sources in one bin is
 $
p_0 = e^{-\varphi(\mu)V_{\rm max}(\mu)}.
 $
A 95 per cent confidence upper limit to $\varphi(\mu)$ can now be expressed   
as
 $
 \varphi(\mu)=- \ln(0.05)/ V_{\rm max}(\mu).
 $
 This equation is used for the upper limits that are indicated by arrows
in Fig.~\ref{sbhist.fig}.

This absence of extreme LSB galaxies suggests two things.  First, the
space density of massive, gas-rich, extremely LSB disks such as Malin~1
(Bothun \etal\ 1987) must be low, as previously shown by 
e.g., Briggs (1990, 1997), Driver \& Cross (2000) and Blanton et al. (2001).
Such a disk would have been easily
detectable to the limit of 7500 \kms\ of the \ahiss, so they must be
intrinsically less common on average than 1 per $1000\, h_{100}^{-3}\,
\rm Mpc^3$ (95 per cent confidence level, using the Poisson statistics).
Second, optical surveys for LSB galaxies (Sprayberry \etal\ 1996;
Dalcanton \etal\ 1997b; O'Neil et al.  1997) systematically find
galaxies
at lower surface brightnesses than $\mu(0) = 24$ \magsq, so there must
be some
reason why the present \hi\ survey fails to detect any.
  
One possibility is that such galaxies have detectable amounts of neutral
hydrogen but are extremely rare, so that it would not be expected to
find one in the \ahiss\ search volume.  Apart from a few special objects
like Malin~1, this seems unlikely because optical surveys find these
objects in significant numbers despite the relatively small volume
limits imposed by optical surface brightness selection effects (McGaugh 
1996).  Specifically, Dalcanton \etal\ (1997b) find that the number
density of galaxies with $V$-band central surface brightnesses in the
range $23<\mu<25~\magsq$ is $0.08 ^{+0.05}_{-0.04}~\ihmpcc$ comparable
to the number density of normal galaxies.  For reference, we have
indicated this estimated with a shaded box in Fig.~\ref{sbhist.fig}, 
where we have adopted $B-V=0.5$, a typical value for LSB galaxies (de
Blok, van der Hulst \& Bothun 1995).

The other possibility is that a significant number of galaxies exist in
the \ahiss\ search volume with optical surface brightnesses $\mu(0) >
24~\magsq$, but that they do not contain enough \hi\ to be detected by
the
\ahiss.  This seems considerably more likely, as there are two ways such
a population could come to exist: First, these very low density systems
could have formed a first generation of stars and then either lost most 
of their remaining gas through supernova-driven winds (Babul \& Rees
1992; Babul \& Ferguson 1996) or consumed all their gas in vigorous
star formation and since then faded to become LSB disks (Bell et al.
1999).  Second, like the outskirts of normal spiral galaxies, LSB disks
have low \hi\ surface densities (de Blok, McGaugh \& van der Hulst
1996), and as such, they are subject to ionization by the extragalactic
UV background that produces the sharp cutoffs seen at the edges of   
normal spirals, for column densities below $10^{19.5}\, \icmsq$ (e.g.,
Maloney 1993; Corbelli \& Salpeter 1994; Dove \& Shull 1994).  Thus,
much of the gas in LSB disks should become ionised, and thus be
undetectable in 21cm surveys.  
This is consistent with the finding that no \ahiss\ galaxies were found
with average \hi\ column densities lower than $\langle N_{\rm HI}
\rangle > 10^{19.7}\,\icmsq$ (Zwaan et al.  1997).
The average limiting column density of the \ahiss\ was $\approx 10^{18}
\rm cm^{-2}$ ($5\sigma$) per 16~\kms.  Of course, for many galaxies in  
the \ahiss\ sample this number is not the minimal detectable \hi\ column
density averaged over the gas disk.  Not all detected galaxies fill the
beam of the Arecibo Telescope, and the velocity width of all detections
is larger than 16~\kms.  A typical \ahiss\ galaxy fills 50 per cent of
the beam and has a velocity width of 160~\kms.  With optimal smoothing
applied, the minimal detectable column density of such a galaxy would be
$10^{18.8}\,\icmsq$, still approximately an order of magnitude lower
than the reported cut-off at $10^{19.7}\,\icmsq$ in Zwaan et al. (1997).
Moreover, the
detection limit of very large galaxies or gas clouds that do fill the
beam and have a similar velocity width of 160~\kms\ would be even lower,
approximately $10^{18.5}\,\icmsq$.  None of these extended, low column
density systems were detected.

At present there is insufficient data to distinguish between the two
proposed hypotheses.  Currently available studies of the stellar compositions of
LSB galaxies (McGaugh \& Bothun 1994; de Blok et al.  1995; Bell et al. 
1999 and 2000) have concluded that gas-rich LSBs form stars slowly and
continuously and therefore have fairly young stellar populations. 
Judging from their colours, the newly identified class of red LSB
galaxies (O'Neil, Bothun \& Schombert 2000) are consistent with a
scenario in which they are simply a fading, passively evolving
population (Bell et al.  1999).  However, O'Neil et al.  (2000) report
that some of these galaxies have high values of $\mhi/L$, but they also
note that 60 per cent of their LSB galaxy sample that was followed-up
with Arecibo is {\em undetected} in \hi, in sharp contrast to the high
success rate in earlier LSB samples (e.g., Schneider et al.  1990). 
Moreover, a cross-correlation of the tables in O'Neil et al.  (1997) and
O'Neil et al.  (2000) shows that the global $V-I$ colours of the
undetected galaxies are on average 0.3 mag redder than the galaxies in
which \hi\ was found. 

Multicolour photometry of galaxies with $\mu_{B} > 24.0~\magsq$, in
combination with deep H-$\alpha$ imaging and deep 21cm observations
should show whether the lowest surface brightness galaxies are
consistent with a fading, passively evolving population. 

\subsection{The LSB contribution to the neutral gas density}
 To address the problem of the cosmological significance of gas-rich LSB
galaxies in a meaningful way, an LSB galaxy should be well-defined.  In
the literature, different authors adopt different definitions for the
critical surface brightness that separates galaxies into the 'normal'
and LSB classes.  The critical value ranges from 21.65~\magsq\ (the
`Freeman value') to 23.5~\magsq.  In the remainder of this paper we
define an LSB galaxy as a galaxy with de-projected $B$-band central
surface brightness $> 23.0~\magsq$.  This limits the LSB galaxies to
those that are $\sim 4\sigma$ dimmer than the Freeman value. 

\begin{figure}
\epsfxsize=8.8cm \epsfbox[10 195 400 670]{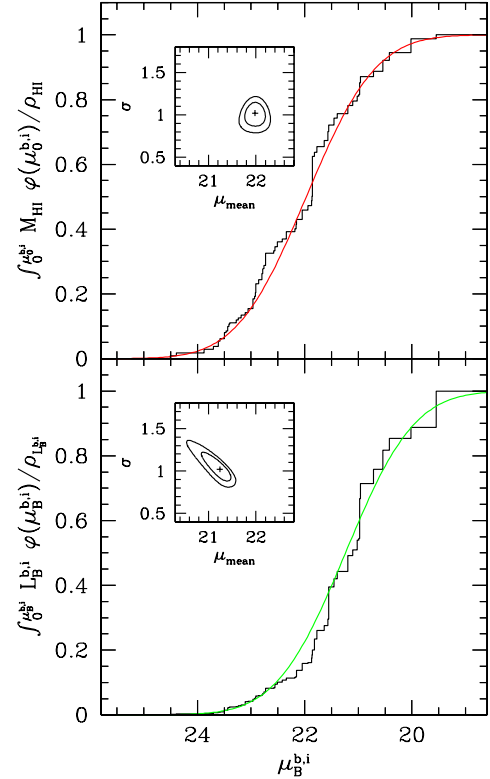}
 \caption{Cumulative distributions of \hi\ mass density (top panel) and
luminosity density (bottom panel) as a function of central surface
brightness for the \ahiss\ sample.  The lines show Gaussian fits.  The
1$\sigma$ confidence levels on the two jointly fitted parameters $\mu$
and $\sigma$, the mean and the width of the Gaussian, are shown in the
inset.  The \hi\ density of the local Universe is dominated by galaxies
with $B$-band central surface brightness of 22.0~\magsq, the luminosity
density is dominated by 21.2~\magsq\ galaxies for this \hi\ selected 
sample.} \label{cumu.fig} \end{figure}

 The cumulative distribution of \hi\ density among \ahiss\ galaxies of
different
surface brightness is presented in the top panel of
Fig.~\ref{cumu.fig}.  The \hi\ density distribution can be fit
satisfactorily with a Gaussian distribution.  There is no fundamental
physical motivation for using a Gaussian to parameterise the
distribution
function, but Dalcanton, Spergel \& Summers  (1997a) note that a galaxy
formation
scenario based on a log-normal distribution of the spin parameter
$\lambda$, produces a (nearly) Gaussian function of luminosity density
vs.  surface brightness (see also de Jong \& Lacey 2000).  The inset in
Fig.~\ref{cumu.fig} gives $1\sigma$ and $2\sigma$ confidence ellipses
for the Gaussian fits, the horizontal axis shows the centre of the      
distribution, the vertical axis the dispersion ($1\sigma$). The actual 
fitting was done on the binned data, not on the cumulative distribution.

The \hi\ mass density of the local Universe is dominated by galaxies
dimmer than the Freeman (1970) value of 21.7~\magsq.  The peak of the
differential
distribution is at 22.0~\magsq, the width is 1.0~\magsq.  Low surface
brightness galaxies contribute a minor fraction to the \hi\ density,
galaxies fainter than 23.0~\magsq\ make up $18_{-5}^{+6}$ per cent of
the \hi\ mass density in the local Universe (the quoted errors have been
determined using bootstrap re-sampling and mark the 68 per cent confidence
levels). 

\subsection{The LSB contribution to the luminosity density}
\label{lumdens.sec}
 The lower panel of Fig.~\ref{cumu.fig} shows the cumulative
distribution of luminosity density against surface brightnesses.  The
peak of the 
differential
distribution is at 21.2~\magsq, the 1$\sigma$ dispersion is
$1.0~\magsq$.  This implies that while most of the \hi\ resides in
galaxies dimmer than the Freeman value, most of the light in gas-holding
galaxies in the local Universe is in galaxies 0.5 mag brighter than the
Freeman value.  The contribution of LSB galaxies is insignificant;
galaxies with $\mubi>23~\magsq$ constitute no more than $5\pm 2$ per
cent to the luminosity density. 

We stress that this result only holds for gas-rich LSB galaxies.  The
contribution of gas-free LSB galaxies is unconstrained by our survey.  
Sprayberry et al.  (1997) conclude that optically selected LSB galaxies
contribute about 30 per cent to the field galaxy luminosity density,  
a result
very consistent with ours, since their definition of an LSB galaxy is
$\mu_B>22.0~\magsq$.  De Jong \& Lacey (2000) find that the luminosity
density
of optically selected galaxies is dominated by $\mu_{\rm I}\sim
19.3~\magsq$, which compares to 21.0~\magsq\ in the $B$-band (using
their
value of $B-I=1.7$).  They estimate that approximately 4 per cent of the
luminosity density is provided by galaxies with $\mubi>22.75~\magsq$.
Driver (1999) defines a volume limited subsample of 47 galaxies at
$0.3<z<0.5$ from the Hubble Deep Field and derives that LSB galaxies 
(mean surface brightness within the effective radius $>21.7~\magsq$)   
contribute $7\pm 4$ per cent to the luminosity density.  All these results are
in good agreement with our estimates.  This fact implies that if a 
population of very LSB, gas-free LSB galaxies exists (as was discussed 
in Section~\ref{cutoff.sec}), their contribution
to the luminosity density must be negligible.

\subsection{The LSB contribution to $\Omega_{\rm matter}$}
 The contribution of LSB galaxies to the total mass budget of the local
Universe is a longstanding question.  The calculation critically depends
on the assumptions one makes on the dependence of the dynamical \ml\ on
central surface brightness.  A zeroth order approximation is to assume 
that \ml\ is equal for all galaxies, independent of central surface 
brightness.  This assumption follows naturally from the observation that
surface brightness is not a parameter in the Tully-Fisher relation
(Sprayberry et al. 1995; Zwaan et al. 1995; Verheijen 1997).  Moreover, 
Verheijen (1997) shows that it is possible to use one model for the dark
matter halo to fit the rotation curves of three galaxies, all at equal
position in the Tully-Fisher relation, but with different surface 
brightness.  In the terminology of McGaugh \& de Blok (1998) this    
invariant \ml\ would be the `same halo hypothesis.'  It is consistent   
with the idea that all galaxies of equal luminosity form in the same   
mass halo, but the angular momentum of an LSB disk is higher, which
causes the disk to be less centrally concentrated (Dalcanton et al.
1997a).
Combined with the result on the luminosity density from
section~\ref{lumdens.sec}, this assumption leads to the conclusion that
$\rho_{\rm M}(\rm LSB)$ is 5 per cent of the total \rhom\ (i.e., equal to the
contribution to the luminosity density).

Van den Bosch \& Dalcanton (2000) show that their semi-analytical galaxy
models are consistent with $\ml \propto \Sigma^{-1/2}$ (Zwaan et al. 
1995), where $\Sigma$ is the central surface brightness in linear units. 
\ml\ ratios are calculated via $M\propto D V^2$, where $V$ is the
maximum rotational velocity and $D$ is a characteristic size of the dark
halo, which is assumed to be directly proportional to the scale length
of the optical disk.  If we, like Driver (1999), adopt this relation for
\ml\, we find that the LSB contribution to \rhom\ rises to
$11_{-3}^{+4}$ per cent.  If we apply the calculations of $M\propto D V^2$
directly to our \ahiss\ data set, we find $\rho_{\rm M} (\rm LSB) =
10_{-3}^{+4}$ per cent.  Both values are in excellent agreement with the $12\pm
6$ per cent that Driver finds. 

At present it is unclear what the true dependence of the dynamical \ml\
on optical surface brightness is.  Clearly, high precision measurements
of rotation curves of LSB systems are needed (see Swaters, Madore \&
Trewhella 2000; van den Bosch \& Swaters 2000; van den Bosch et al. 
2000).  At the moment we adopt as a conservative estimate that gas-rich
LSB galaxies contribute no more than 11 per cent to \rhom, the dynamical mass
contained in galaxies.

\section{Bivariate distributions} \label{bivar.sec}
 A more detailed view of the distribution of baryons among galaxies of
different size and brightness can be obtained by calculating bivariate
distribution functions.  The importance of this way of looking at galaxy
parameters is stressed by van der Kruit (1987, 1989), de Jong (1996),
and most recently by de Jong \& Lacey (2000) who study $10^3$ galaxies
with types Sb to Sdm.  
 First, the bivariate distribution function is a important
constraint for galaxy formation theories, as any theory should not only
produce the integrated luminosity function (and integrated distribution
functions of other structural parameters), but also higher dimensional
distribution functions.  Second, bivariate distribution functions help
to clarify the selection effects that influence the determination of
(e.g.) the luminosity function. 

The aim of the present work is 1) to test whether an \hi\ selected
galaxy sample yields qualitatively the same bivariate distribution
function, and, 2) extend the bivariate distribution functions to the
distribution of \hi\ properties.  The sample we study here is small,
and, as is discussed by de Jong \& Lacey (1999) and Minchin (1999), at
least $500-1000$ galaxies are required to avoid problems with small
number statistics.  We only intend to make qualitative comparisons and
care should be taken with the interpretation of the results.  We also
stress that the sample that we use here might be biased against galaxies
with very low values of $\mhi/L$, just like optical samples are biased
to those with high values of $\mhi/L$.

\subsection{Results}
\begin{figure*}
\epsfxsize=14cm \epsfbox[18 144 592 718]{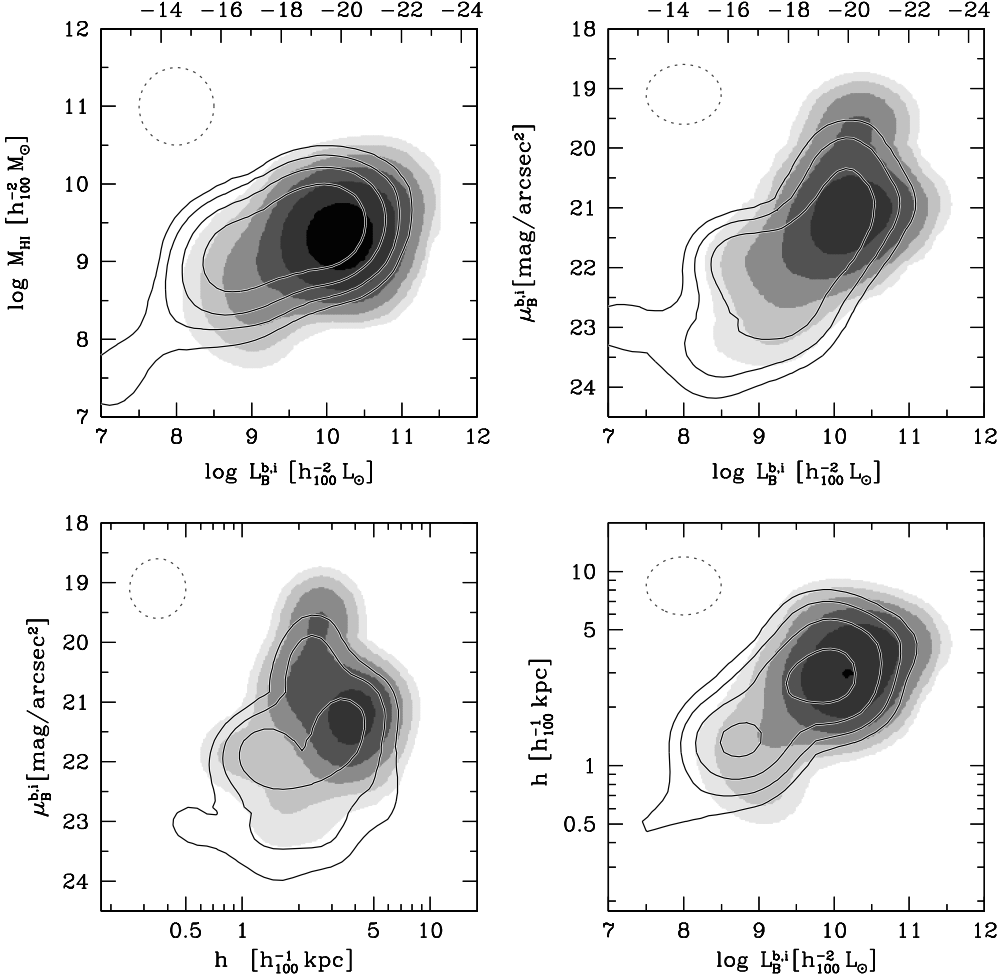}
 \caption{Bivariate distribution of luminosity density (grey-scales) and
\hi\ density (contours) for \hi\ selected galaxies in
 the (\lbi,  \mhi)-plane  ({\em top   left\/}),
 the (\lbi, \mubi)-plane  ({\em top   right\/}),
 the ($h$,  \mubi)-plane  ({\em lower left\/}), and
 the (\lbi,   $h$)-plane  ({\em lower right\/}).
 Grey-scales correspond to $(10^{6.5}, 10^{6.75}, 10^{7.0},...)\times
       h_{100}\, L_\odot\,\rm Mpc^{-3}$,
 contours to $(10^{6.25}, 10^{6.5}, 10^{6.75},...)\times
h_{100}\, M_\odot\,\rm Mpc^{-3}$. The densities are per decade for \lbi\
and \mhi, per 0.3 dex for $h$ and per mag for \mubi. The dashed ellipses
in the upper right corners of each panel indicate the FWHM of the   
Gaussian smoothing filter that has been applied to the data.}
\label{bivar.fig} 
\end{figure*}

 In Fig.~\ref{bivar.fig} we present bivariate distributions of several
fundamental parameters.  Our aim is to show the distribution of \hi\
mass density and luminosity density as a function of galaxy luminosity, 
gas mass, size, and surface brightness.

The figures are calculated by distributing values of $\mhi/\vmax$ and
$\lbi/\vmax$ over a fine grid with 0.1 dex resolution for \mhi\ and
\lbi, 0.1 mag resolution for central surface brightness
\mubi, and 0.05 dex resolution for disk scale length $h$.
Next, the images were smoothed with a Gaussian filter, of which the FWHM
values are indicated by the dashed ellipses in the upper left corners of
each panel.  The \hi\ density distributions are shown as contours, the
luminosity density distribution as grey-scales.  Steps in intensity are 
in logarithmic intervals of 0.25 dex.

The first thing to notice is that the general trends are the same for
the gas density and the luminosity density, but the maximum of the gas
density is shifted towards less luminous, lower surface brightness
galaxies.  The second point is that the luminosity density is more
strongly concentrated towards large, luminous HSB galaxies, whereas the
\hi\ density is more widely distributed.

What is obvious from the top right panel is that both the luminosity and
the \hi\ mass distribution are strongly dependent on optical surface
brightness in the sense that both functions are shifted towards fainter
absolute magnitudes for lower surface brightness galaxies.  This fact
was also observed by de Jong (1996), and the peak of the distribution,
at $\mbi=-20+5 \log h_{100}$ and $\mubi=21~\magsq$ agrees well with his
determination.  The same trend is observed by Cross et al.  (2001) who
present a bivariate distribution function based on a a preliminary
subsample of $5\times 10^4$ galaxies from the 2dF survey and by Blanton
et al.  (2001) who use a sample of $10^4$ galaxies from the SDSS
commissioning data.

A similar effect can be seen in the lower right panel:
the luminosity and the \hi\ mass distribution are shifted towards
fainter absolute magnitudes for smaller galaxies.  This correlation has 
been studied in detail by de Jong \& Lacey (2000), who discuss the
predictions of hierarchical galaxy formation theories and conclude that
the observed distribution is in qualitative agreement with theory, but
the distribution in disk size is narrower than predicted.

\section{Conclusions} \label{conc_opt.sec}
 We have presented several volume corrected distribution functions for
galaxies that have been selected in the \hi\ 21cm line. The conclusions
are the following:

1. The luminosity function of the \hi\ selected galaxies is in agreement
with other determinations based on late-type, or star-forming galaxies.
The integral luminosity density of gas-rich, late-type, or star-forming
galaxies is well determined and equals $\rhol=(3.4\pm 0.7) \times
10^{19}\,h^{-2}_{100}\rm\,W\,Hz^{-1}\,Mpc^{-3}$.  This is approximately
50 per cent of the integral luminosity density of the local Universe.

2. The contribution of low surface brightness (LSB) galaxies to the
integral luminosity density and \hi\ density is modest, 5 and 18 per cent,  
respectively.  This in in good agreement with calculation based on
optically selected galaxies. The fraction of $\Omega_{\rm matter}$ that
resides in LSB galaxies is at present not well determined, but probably
less than 11 per cent.

3.  We observe a lower limit to the surface brightness of gas-rich
galaxies: no galaxies were found with de-projected central surface
brightness $> 24.0~\magsq$ in the $B$-band.  It will be interesting to
test whether this result stands up with future large \hi\ surveys, such
as the HIPASS survey (Staveley-Smith et al.  1996). 

4. Bivariate distributions of various fundamental galaxy parameters show
that the \hi\ density in the local Universe is more diffusely spread   
over galaxies with different size, surface brightness, and luminosity  
than the luminosity density.  The luminosity density is concentrated
towards bright, large, high surface brightness disks.

\section*{Acknowledgements}
We thank L. Staveley-Smith, R. Webster and B. Koribalski for helpful
comments. 
 The Isaac Newton Telescope is operated by the Royal Greenwich
Observatory in the Spanish Observatorio del Roque de los Muchachos of
the Instituto de Astrofisica de Canarias.  The National Radio Astronomy
Observatory is a facility of the National Science Foundation operated
under cooperative agreement by Associated Universities, Inc.  The
Arecibo Observatory is part of the National Astronomy and Ionosphere
Center, which is operated by Cornell University under a cooperative
agreement with the National Science Foundation.  The National Radio
Astronomy Observatory is a facility of the National Science Foundation
operated under cooperative agreement by Associated Universities, Inc.

\label{lastpage}

\end{document}